\documentclass[amsmath,amssymb,aps,superscriptaddress]{revtex4-2}

\usepackage{graphicx}
\usepackage{dcolumn}
\usepackage{bm}
\usepackage{hyperref}
\hypersetup{
    colorlinks = true,
    urlcolor   = blue,
    citecolor  = blue,
    linkcolor  = blue,
}
\usepackage{lineno}
\usepackage{textcomp}
\usepackage{amsmath}

\begin{document}
\preprint{APS/123-QED}

\title{Viscous influences on impulsively generated focused jets}

\author{Xianggang Cheng}
\affiliation{School of Marine Science and Technology, Northwestern Polytechnical University, Xi’an 710072, China}
 
\author{Xiao-Peng Chen}
\email{xchen76@nwpu.edu.cn}
\affiliation{Research \& Development Institute of Northwestern Polytechnical University in Shenzhen, Shenzhen 518057, China}
\affiliation{School of Marine Science and Technology, Northwestern Polytechnical University, Xi’an 710072, China}
\author{Hang Ding}
\affiliation{Department of Modern Mechanics, University of Science and Technology of China, Hefei 230027, China}
\author{Chun-Yu Zhang}
\affiliation{Department of Modern Mechanics, University of Science and Technology of China, Hefei 230027, China}
\author{Haibao Hu}
\affiliation{School of Marine Science and Technology, Northwestern Polytechnical University, Xi’an 710072, China}

\author{Laibing Jia}
\email{l.jia@strath.ac.uk}
\affiliation{Department of Naval Architecture, Ocean \& Marine Engineering, University of Strathclyde, Glasgow G4 0LZ, UK}

\date{\today}

\begin{abstract}
Impulsively generated focused jets play a significant role in various applications, including inkjet printing, needle-free drug delivery, and microfluidic devices. As the demand for generating jets and droplets from medium- to highly viscous liquids increases, understanding the role of viscosity in jetting dynamics becomes crucial. While previous studies have examined the viscous effects on walls, the impact on free surfaces has not been thoroughly understood. This study aims to bridge this gap by integrating experiments with numerical simulations to investigate the viscous effects on focused jet formation. We demonstrate that mass and momentum transfer along the tangential direction of the free surface contribute to focused jet formation, and viscosity plays a key role in this transfer process. The viscosity-induced diffusion of the shear flow and vorticity near the free surface reduces the jet speed. Based on experimental observations and simulation results, we propose an equation to predict the viscous jet velocity. These findings offer new perspectives on viscous interface dynamics in advanced manufacturing and biomedical applications.
\end{abstract}

\maketitle

Liquid jets are a fundamental phenomenon in both natural settings and industrial applications\cite{Eggers2008}. They are widely observed on liquid surfaces, from the bursting bubbles in champagne that enhance sensory experiences\cite{Liger2021}, to the splashing of ocean waves that impact the global climate\cite{Brooks2018,Deike2022}, and even the transmission of pollutants and pathogens through contaminated jets that can harm human health\cite{Bourouiba2021,Chen2024}. At the core of these diverse phenomena lies the physical process of mass and momentum transfer across air-liquid interfaces. Building on the exploration of the underlying physics, liquid jets are utilised in a broad range of technological advancements and industrial processes, such as inkjet printing\cite{Modak2020,Lohse2022}, needle-free drug delivery\cite{Rohilla2020,Schoppink2022}, and various other applications.

The formation of liquid jets is primarily initiated by rapid changes in pressure, which occur during the collapse of an air bubble floating on the water's surface\cite{Deike2018,Ji2023}, the oscillation of a cavitation bubble beneath an air-liquid interface\cite{Rossello2022}, or the impact of a droplet on a solid or liquid surface\cite{Lin2021,Tian2023}. To replicate these rapid pressure variations that lead to jet formation in research, one effective method involves the generation of a highly focused jet from a concave meniscus. This can be achieved either by suddenly impacting the liquid container\cite{Antkowiak2007,Kiyama2016,Zhang2020,Cheng2021} or by triggering an explosive vapour bubble beneath the surface\cite{Tagawa2012,Zhang2017,Rohilla2023}.

In these experimental methods, the generation of a focused jet can be divided into an impact stage ($t\le$$10^{-4}~\rm{s}$) and a flow-focusing stage ($t\gg$$10^{-4}~\rm{s}$)\cite{Peters2013,Gordillo2020}. During the impact period, the sudden impact induces a pressure impulse, which in turn generates the initial velocity inside the liquid bulk\cite{Cooker1995,Antkowiak2007}. Based on the pressure impulse theory, Antkowiak et al.\cite{Antkowiak2007} provided an analytical solution for the velocity and pressure field just after the impact. In the subsequent flow-focusing stage, the liquid surface evolves freely and ejects a focused jet. The flow-focusing effect of normal liquid velocities at the locally spherical meniscus counts for the velocity increment, and further leads to the occurrence of a jet\cite{Peters2013,Gordillo2020}. Moreover, the jet velocity $V_j$ can be theoretically predicted as $V_j \propto U_0$, where $U_0$ is the interface velocity just after impact.

Existing theoretical predictions based on low-viscous or inviscid liquids have shown good agreement with experimental and numerical results\cite{Peters2013,Gordillo2020}. However, with the increasing demand for generating medium- to highly viscous jets and droplets in applications, such as biomaterials printing, spray coating, and electronics manufacturing, the viscous effects in jet dynamics become crucial\cite{Gudapati2016,Turkoz2018PRL,Buchner2023}.
Tagawa et al.\cite{Onuki2018,Kamamoto2021} extended the study of focused jets to a wider range of liquid viscosities and found that higher liquid viscosities decreased jet velocity. Numerical simulations confirmed that the viscous effect is negligible during the impact stage. In contrast, viscosity plays an important role during the flow-focusing stage, and it is further considered that the development of the wall boundary layer due to viscosity interrupts flow-focusing.
Moreover, there have been a few studies focusing on the dynamics of jets on oil-covered water surfaces, which arise from the presence of an organic microlayer or oil spill on the sea surface\cite{Ji2021,Cheng2021,Yang2023}. These studies show that even with a thin oil layer, the jet tip radius and velocity are significantly altered, suggesting that oil spreading on the water surface influences the effective viscous damping and, consequently, the jetting dynamics. These results imply that the flow near the free surface also plays a role in jet dynamics and should be considered when studying viscous jet formation.

However, the viscous effect on the free surface dynamics has not been thoroughly understood\cite{Riquier2024}. To address this gap, we conducted tube-impact experiments on focused jets using liquids with different viscosities. Numerical simulations based on the diffuse interface method\cite{Zhang2016,Yang2018,Guo2020POF,Guo2020EML,Zhang2021,Zhang2023} were also performed to gain more insight into the flow details. 
The aim of this study is to investigate the viscous effects on the free surface during focused jet formation process and understand their impacts on jetting dynamics.

\begin{figure*}[t!]
\includegraphics[width=1\textwidth]{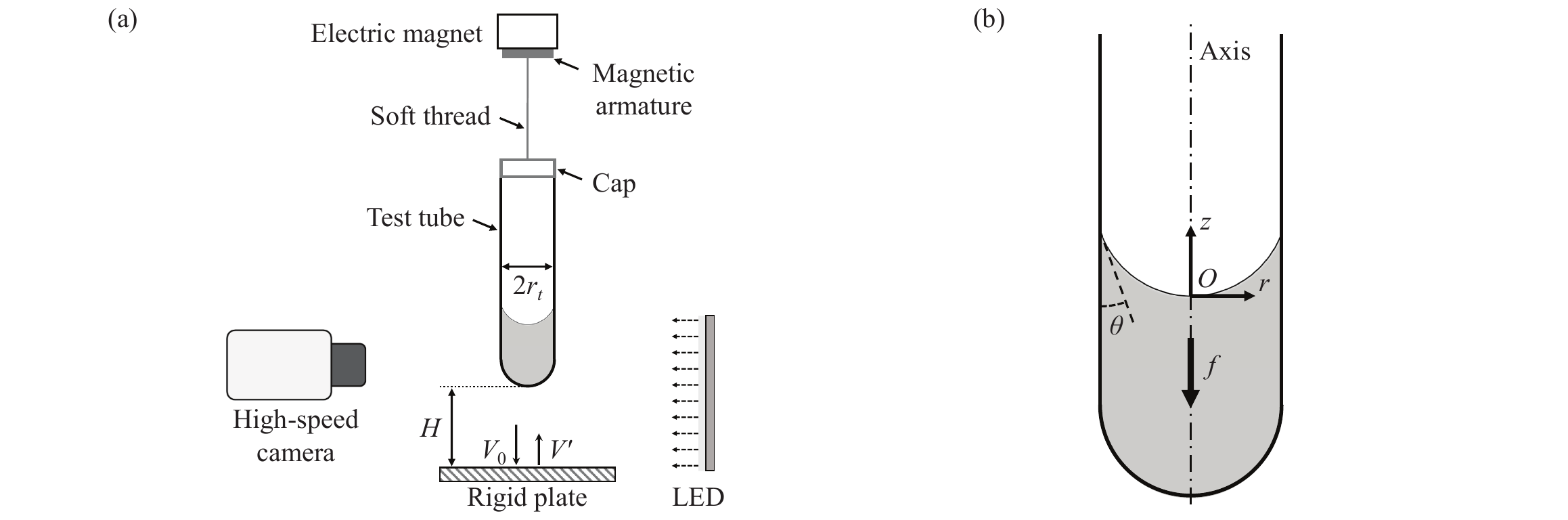}
\caption{Sketches of the experimental and numerical set-up. (a) Experimental setup. A test tube was initially held by an electromagnet to a height $H$. $V_0$ and $V'$ denote the impact and rebound speed of the test tube just before and after impact, respectively. (b) A schematic of the axisymmetric numerical set-up. The grey region denotes the liquid phase, and the contact angle with the wall was set at $\theta=20^\circ$ based on the experimental measurement. An impulsive force $f$ was applied in the liquid region to simulate the impact. 
\label{Fig.Setup}}
\end{figure*}

The experimental configuration is shown in Fig.~\ref{Fig.Setup}(a). A test tube was partially filled with a working liquid to a depth of $L=7$ mm. The test tube is made of glass, with an inner radius of $r_t=4.0$\,mm, a length of 100 mm, and a wall thickness of 1.0 mm. The mass of each tube is $6.9 \pm 0.2$ g (excluding the cap). The test tube cap is made of polylactide (PLA) using a 3D printer (Replicator 2, MakerBot). The cap has an inner radius of 5.0 mm, a height of 10 mm, and a wall thickness of 1.0 mm, with a mass of approximately 0.5 g. A magnetic armature is attached to the tube cap by a soft thread.
The tube was held by an electromagnet at a certain height $H$ from a rigid plate underneath. The rigid plate is made of stainless steel, with dimensions of $100 \times 100 \times 5~\mathrm{mm}^3$.
When the experiment started, the magnet was switched off, and the tube fell freely until it impacted the rigid plate. The test tube then rebounded, and a focused jet was ejected from the liquid surface. A high-speed camera (Megaspeed-75K, MegaSpeed Corp., Canada) equipped with parallel back lighting was employed to capture the jetting process. The frame rate was set at 5,000 fps with an exposure time of 100~$\mu$s. In the experiments, each working condition was repeated at least three times for reproducibility.

\begin{figure*}[t!]
\includegraphics[width=0.9\textwidth]{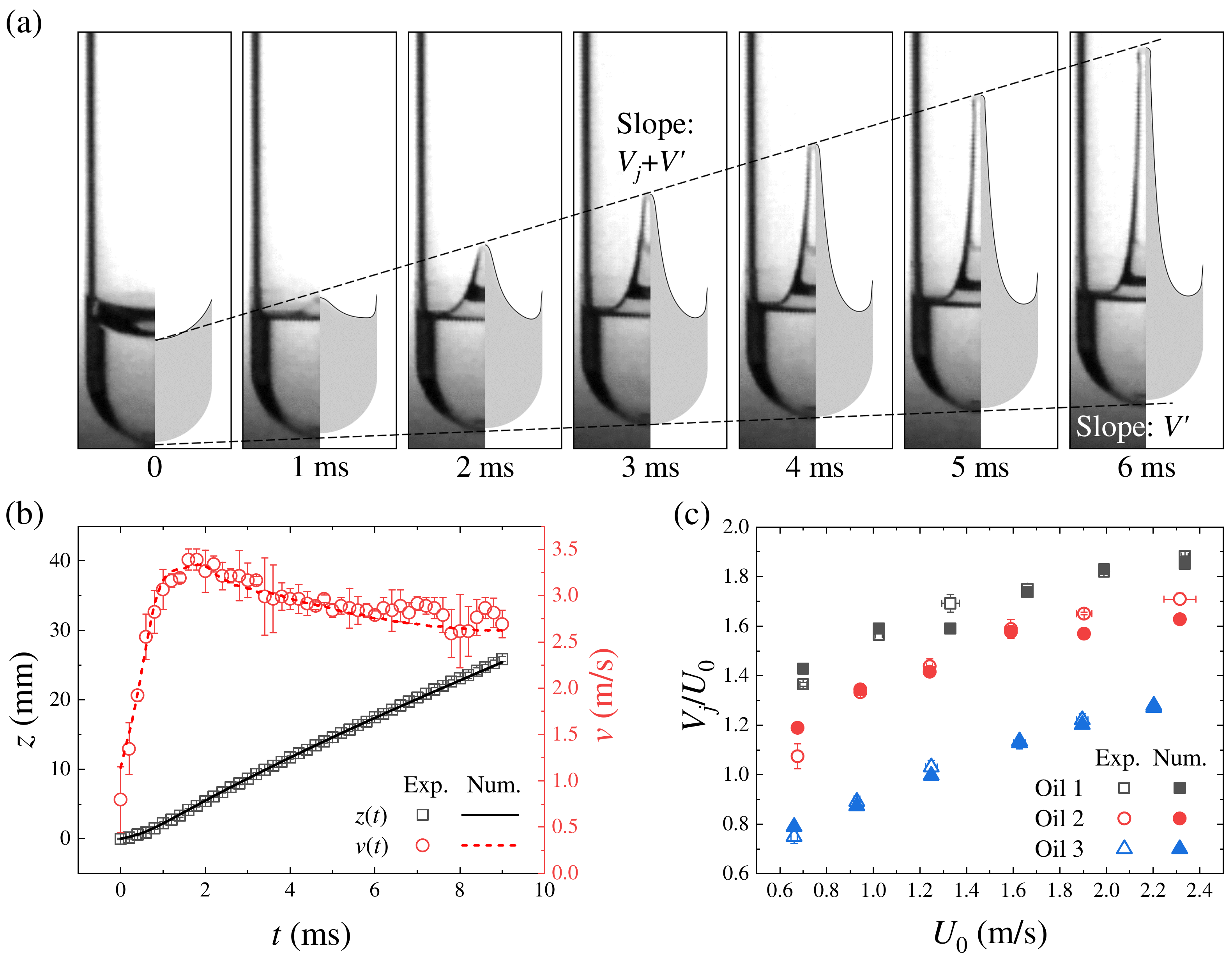}
\caption{Comparison between experimental and numerical results. (a) Temporal evolution of a typical jetting process. The left of each panel shows the experimental images. Here Oil-1 was used and $U_0=1.66~{\rm m/s}$. The right of each panel shows the numerical snapshots of the corresponding interface shape. The black curves denote air-liquid interfaces. The dashed lines represent the positions of the jet tip and the test tube bottom, and their slopes are velocities $V_j + V'$ and $V'$, respectively. (b) The jet tip position ($z(t)$) and the instantaneous interface velocity ($v(t)$) in the coordinates fixed on the test tube, corresponding to the case in (a). The experimental data (scatters) are the average results of three runs of experiments under the same working condition, and the error bars are plotted. The lines are the corresponding numerical results. (c) Dimensionless jet velocities $V_j/U_0$ under various impact velocities $U_0$ and working liquids. Hollow symbols and solid ones denote experimental and numerical data, respectively.\label{Fig.Validation}}
\end{figure*}

A characteristic impact velocity $U_0$ was defined as,
\begin{equation}
 U_0=V_0+V',   
\end{equation}
where $V_0$ and $V'$ denote the impact and rebound speed of the test tube just before and after impact, respectively (Fig.~\ref{Fig.Setup}a). $V_0$ can be theoretically determined as $\sqrt{2gH}$, which was further validated by the experimental measurement. $g$ denotes the gravitational acceleration, $g=9.8~{\rm m/s^2}$. $V'$ is the average speed of the test tube over a time period of $\Delta t=6~{\rm ms}$ after impact (Fig.~\ref{Fig.Validation}a). In our experiment, $U_0$ ranged from $0.7~{\rm m/s}$ to $2.3~{\rm m/s}$, corresponding to falling heights $H=15\sim 130~{\rm mm}$. 
The jet velocity $V_j$ is the average interface velocity in the frame fixed to the test tube, over a time period of $\Delta t = 6~\mathrm{ms}$ after impact, reflecting a broader temporal scope (Fig.~\ref{Fig.Validation}a).
Given the characteristic duration for jet formation\cite{Peters2013}, $t_c \sim r_t/U_0$, the maximum $t_c$ in our experiments is approximately 5.7 ms with $U_0 = 0.7$~m/s. For higher $U_0$, $t_c$ decreases. Here we selected $\Delta t = 6$ ms for all experiments to ensure the jets were fully developed, as this duration encompasses the entire jet formation period. The value of $\Delta t$ is consistent with the study by Kiyama \textit{et al.}\cite{Kiyama2016}.
Additionally, the averaged jet velocity is normalised by the impact speed, denoted as $\beta \equiv V_j/U_0$.

Three types of silicone oils (PMX-200, Dow Corning) were selected to explore the viscosity effects (Table~\ref{Tab.Prop}). The kinematic viscosities range from $11$ to $57~\mathrm{mm^2/s}$. The viscosities were measured using Vickers viscometers (Loikaw Instrument, China) at $20^\circ \mathrm{C}$. The liquid density was calculated by weighing a certain volume of liquid, and the surface tension was measured using a pendant drop method\cite{Berry2015}.

\begin{table}[htbp]
\centering
\caption{Physical properties of the working liquids.
\label{Tab.Prop}}
\begin{tabular}{c c c c}
\hline 
  & $\rho_l~({\rm kg/m^3})$ & $\sigma~({\rm mN/m})$ & $\nu~({\rm mm^2/s})$ \\ 
\hline 
Oil-1 & 937 & 20 & 11 \\ 
Oil-2 & 934 & 19 & 22 \\ 
Oil-3 & 941 & 20 & 57\\ 
\hline 
\end{tabular}
\end{table}

Numerical simulations were conducted using a commercial software COMSOL Multiphysics. An axisymmetric numerical model was established with coordinates fixed to the test tube. As shown in Fig.~\ref{Fig.Setup}(b), the geometry of the numerical model is the same as that in the experiments. 
In the experiments, the tube was released from a height of 15 mm or higher. Its falling time, $T_f=\sqrt{2H/g}\geq55$ ms. Once released, two types of waves were generated: capillary waves and waves due to transient free-fall conditions. These waves damped out quickly due to viscosity. Prior to the impact, with 15 ms recorded in the experiments, no noticeable fluctuations of the liquid meniscus were observed. Measurements of the positions of the meniscus bottom and the tube bottom also confirmed that the deformation of the liquid surface during free fall was negligible. The pre-impact liquid surface exhibited a spherical-cap shape in the gravity-free state. In the simulations, the initial shape of the free surface was set as a circular arc determined by $\theta$. By measuring the profiles of the liquid surfaces at the pre-impact moment in the experiments, $\theta = 20^\circ$.

Considering the impact acceleration is much larger than gravity, an effective downward impulsive force $f$ was imposed on the liquid to simulate the impact,
\begin{equation}
    f(t)=ae^{-\frac{(t-\tau/2)^2}{2c^2}},
\end{equation}
where $\tau$ is the time duration of the impact. To match the experiments, $\tau=1~\mu \mathrm{s}$ and $c=\tau/6$ were adopted. The value of $a$ was determined according to the momentum conservation during the impact,

\begin{equation}
\int_0^\tau f(t)dt=\rho_l U_0.
\end{equation}

The diffuse interface method was employed to capture the liquid-gas interface\cite{Zhang2016,Yang2018,Guo2020POF,Guo2020EML,Zhang2021,Zhang2023}. A phase index $\phi$ was defined, where $\phi=1$ in the liquid phase, $\phi=-1$ in the gas phase, and $-1<\phi<1$ across the interface. The interface was updated according to the Cahn-Hilliard equation,
\begin{equation}
\frac{\partial \phi}{\partial t}+\boldsymbol{u} \cdot \nabla \phi=\nabla \cdot(M \nabla G),
\end{equation}
where $\boldsymbol{u}$ denotes the velocity, and $M$ is the mobility parameter that characterises the diffusion rate of the phase. $G$ is the chemical potential, $G=\lambda[-\nabla^{2}\phi+\phi(\phi^{2}-1)/\epsilon^{2}]$, where $\lambda$ is the mixed free energy density, and $\epsilon$ characterises the interface thickness.
The local properties of the material, such as density $\rho$ and dynamic viscosity $\mu$, are interpolated as follows,
\begin{equation}
\begin{split}
    \rho = \frac{1+\phi}{2}\rho_{l} + \frac{1-\phi}{2}\rho_{g},\\
    \mu = \frac{1+\phi}{2}\mu_{l} + \frac{1-\phi}{2}\mu_{g}, 
\end{split}
\end{equation}
where the subscripts $l$ and $g$ denote the liquid and gas phases, respectively.
The surface tension term, $G \nabla \phi$, can be applied in the Navier-Stokes equation,
\begin{equation}
\rho\left(\frac{\partial \boldsymbol{u}}{\partial t}+\boldsymbol{u} \cdot \nabla \boldsymbol{u}\right)=-\nabla p+\nabla \cdot\left[\mu\left(\nabla \boldsymbol{u}+(\nabla \boldsymbol{u})^{\mathrm{T}}\right)\right]+G \nabla \phi+\boldsymbol{f}.
\end{equation}
Furthermore, to obtain accurate results, the interface thickness should satisfy the sharp interface limit\cite{Yue2010}. In our simulations, the mesh size was smaller than 0.01 mm and $\epsilon=0.005~\rm{mm}$, which guaranteed the resolution in capturing the interface.

We compared the jetting results from experiments and simulations in Fig.~\ref{Fig.Validation}. The numerical results are in good agreement with the experimental ones.
Figure~\ref{Fig.Validation}(a) shows typical jetting snapshots, where the left of each panel shows the experimental images and the right shows the corresponding numerical results. Oil-1 is used as the working liquid, and the characteristic impact velocity $U_0$ is $1.66$ m/s. After the impact, the tube rebounds. The liquid surface near the wall sinks, while the surface near the axis rises, forming a jet tip at $t=1$ ms. As the surface continues to evolve, the jet elongates as the tube rebounds over time $t=1\sim6$ ms.
From the snapshots, the temporal evolution of the jet tip position ($z(t)$) and the corresponding instantaneous velocity ($v(t)$) in the coordinates fixed to the test tube were obtained, as shown in Fig.~\ref{Fig.Validation}b. The liquid surface gains an initial velocity immediately after impact, then continues accelerating until $t\approx1.6~{\rm ms}$. Afterwards, deceleration occurs due to surface tension and viscosity ($t>1.6$ ms).
$V_j$ is the average of $v(t)$ over the time $t =0\sim6$ ms.
In our experiments, the Weber number, $We = \rho_l U_{0}^{2} r_t / \sigma$, ranges from 82 to 1054, suggesting a weak influence of surface tension.

The dimensionless jet velocity ($\beta \equiv V_j/U_0$) under various impact velocities and liquid viscosities is summarised in Fig.~\ref{Fig.Validation}(c). The results show that $\beta$ decreases with decreasing $U_0$ or increasing $\nu$.

Considering using $r_t$ as the characteristic length to define the Reynolds number, $Re=r_t U_0/\nu$. $Re$ ranges from 46 to 849 in the experiments. The viscous dissipation rate is roughly proportional to $1/Re$, with estimates of $2\%$ for Oil-3 and 1\textperthousand\ for Oil-1, indicating that the viscous dissipation is negligible. However, this conflicts with the results shown in Fig.~\ref{Fig.Validation}(c), where the Oil-3 jets achieve only about $40\%$ of the kinetic energy ($\propto \beta^2$) compared to Oil-1. This discrepancy suggests that a more detailed analysis of flow structures is needed to understand the viscous effects on jet formation.

\begin{figure*}[t!]
\includegraphics[width=1.0\textwidth]{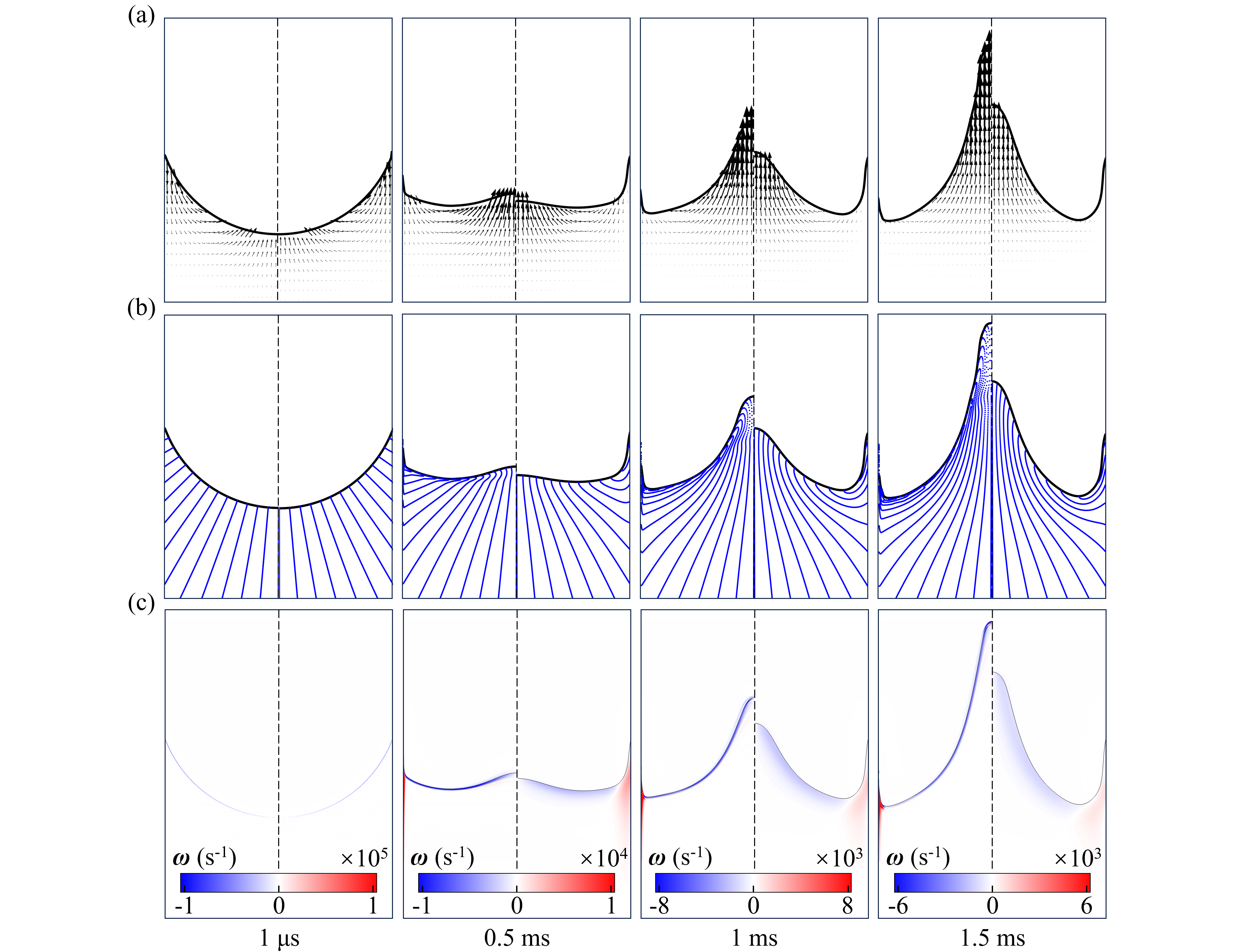}
\caption{Flow details on the focused jet from the numerical results. Here the working liquid is Oil-3 and the impact velocity $U_0$ is $2.30~{\rm m/s}$. For comparison, the results for the nearly inviscid jet ($\nu = 0.5~{\rm mm^2/s}$) are shown on the left of each panel. The solid black curve denotes the air-liquid interface.
(a) Velocity field. (b) Fluid parcel tracing results. The blue lines are made of tracked parcels, which are initially aligned in the normal direction of the surface. (c) Vorticity field.\label{Fig.Vort}}
\end{figure*}

The numerical simulations provide insights into critical aspects of jet formation, such as the velocity and vorticity fields, as well as fluid parcel tracing.
As shown in Fig.~\ref{Fig.Vort}, the right of each panel shows numerical snapshots of Oil-3 with $U_0=2.30$ m/s. For comparison, the results for a nearly inviscid case ($\nu = 0.5~{\rm mm^2/s}$) are shown on the left of each panel with the same $U_0$.
Figure~\ref{Fig.Vort}(a) shows the velocity field in the liquid during jet formation. Over the impact stage, the liquid undergoes an acceleration with the order of $U_0/\tau$, and an initial velocity field ($\sim U_0$) is induced by the pressure gradient ($t=1~{\rm \mu s}$). Moreover, it is noted that the velocity distribution is nearly identical for the viscous and inviscid jets just after impact, suggesting that the viscous effect is negligible during the initial phase\cite{Antkowiak2007,Onuki2018}.

After the flow field is initialised, the liquid surface evolves freely. The liquid near the axis moves upward along it, while the liquid far away from the axis flows along the surface to the bottom of the concave surface, forming a focused flow field.
The trend of focused flow becomes more pronounced at $t=0.5~{\rm ms}$, when the surface is almost flat. As the surface continues to evolve, a jet tip is formed ($t=1~{\rm ms}$), and the jet continues to extend under its own inertia ($t>1~{\rm ms}$). Additionally, the viscous jet gains a lower speed increment during the flow-focusing stage due to the onset of viscous effects.

To enhance clarity, we tracked lines of liquid parcels initially aligned perpendicular to the surface (see Fig.~\ref{Fig.Vort}b, $t=1~{\rm \mu s}$). These parcels illustrate the transportation of liquid elements, showing the mass source of the resulting jet. In both cases, the space between the liquid parcels is stretched in the direction perpendicular to the surface. For the nearly inviscid jet case shown on the left-hand side panels, the parcel lines just beneath the surface bend significantly towards the axis after $t=1~\mu \rm{s}$.
This bending indicates that a substantial portion of the jet's mass originates from the liquid's near-surface region. The large travelling distance of the parcels also suggests that a greater amount of kinetic energy is incorporated into the jet. Conversely, in the case of the viscous jet, the parcel lines undergo a smoother deformation, indicating that the bulk of the jet's mass is drawn from a deeper region below the surface, transferring less kinetic energy into the jet.

The tangential flow near the free surface can be further perceived from the vorticity field (Fig.~\ref{Fig.Vort}c).
Just after impact ($t=1~{\rm \mu s}$), an extremely thin but strong vorticity sheet forms due to the misalignment of density and pressure gradients across the surface, namely $\nabla\rho\times\nabla p\neq 0$~\cite{Lundgren1999,Terrington2020}. 
As the free surface evolves, the vorticity diffuses into the deeper region due to viscous diffusion ($t>1~{\rm \mu s}$). A stronger diffusion homogenises the velocity in the liquid phase, reducing the kinetic energy entering the jet.
For the nearly inviscid jet, the diffusion was weak and the vorticity remained concentrated in a thin layer, resulting in a strong shear flow near the free surface (Fig.~\ref{Fig.Vort}b).
Additionally, the vorticity magnitude is similar across the three oil cases. However, as viscosity increases, the vorticity layer diffuses deeper, and its magnitude decreases.

This phenomenon is related to the development of a viscous boundary layer near the free surface, and its thickness depends on the liquid viscosity and the evolution time\cite{Batchelor1967,Antkowiak2007,Terrington2020}. Given the flow-focusing timescale\cite{Peters2013}, $t_c \sim r_t/U_0$, the boundary layer thickness is estimated as
\begin{equation}
    \delta = \sqrt{\nu t_c} = \sqrt{\frac{\nu r_t}{U_0}}.
    \label{Eq.delta}
\end{equation}

The viscosity-induced diffusion of the shear flow and vorticity near the free surface accounts for the reduction in the jet speed. Utilising $\delta$ as the characteristic length, a new Reynolds number can be defined as
\begin{equation}
    Re_p = \frac{\delta U_0}{\nu} = \sqrt{\frac{r_t U_0}{\nu}} = Re^{\frac{1}{2}}. 
\end{equation}

\begin{figure}[t!]
  \includegraphics[width=0.6\textwidth]{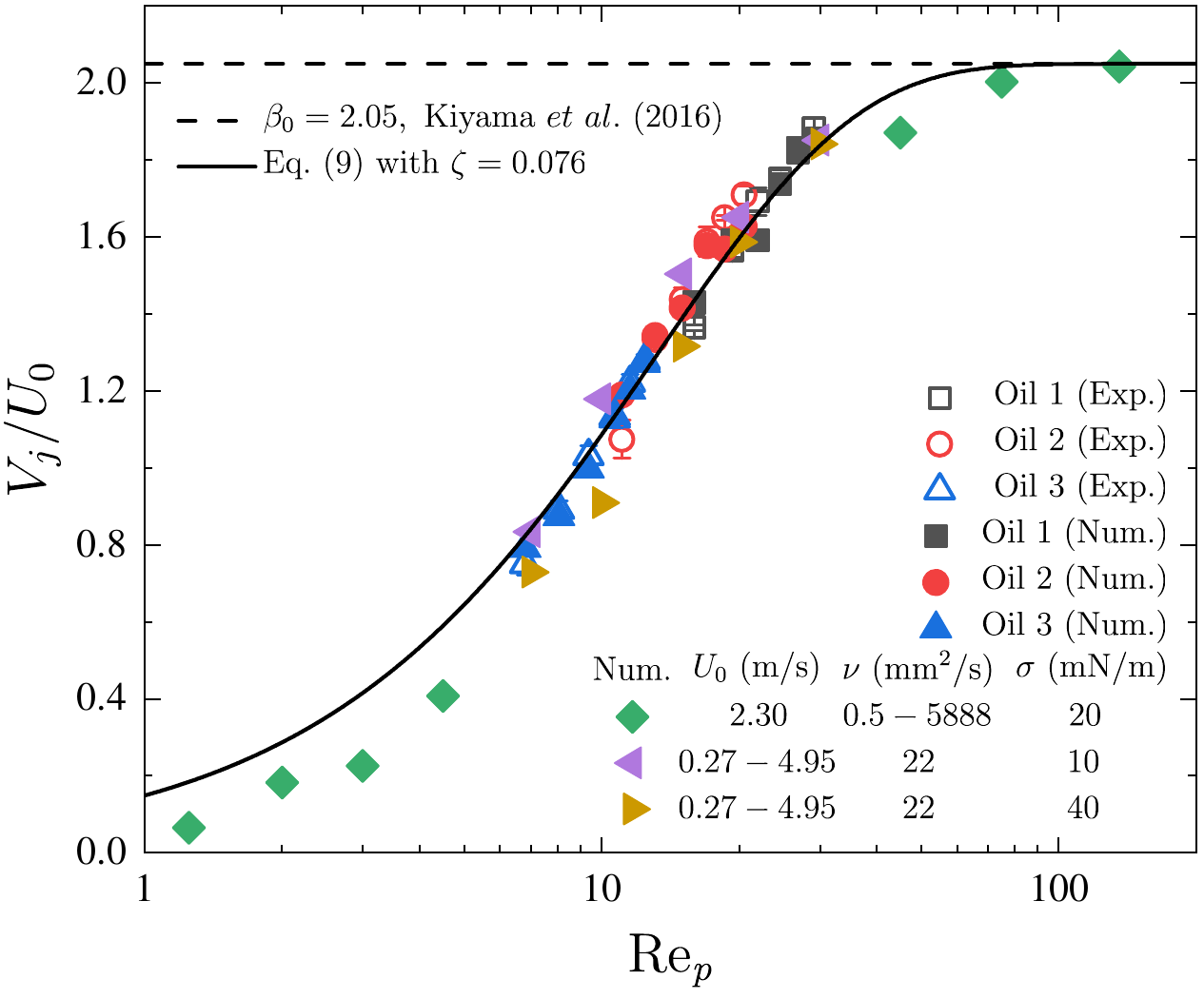}
  \caption{Dimensionless jet velocity $V_j/U_0$ versus the Reynolds number $Re_p$. The scatters are the experimental and numerical data. The solid curve represents the data fitting by Eq.~(\ref{Eq.beta}). The dashed line is the nearly inviscid dimensionless jet velocity, $\beta_0 = 2.05$, obtained from reference \cite{Kiyama2016}.\label{Fig.jetv}}
\end{figure}

The dimensionless jet velocity ($\beta \equiv V_j/U_0$) vs $Re_p$ is plotted in Fig.~\ref{Fig.jetv}, where $Re_p$ ranges from 7 to 29 in our experiments. Additional data with higher and lower $Re_p$ were obtained from numerical simulations with $\nu=0.5\sim5888~\mathrm{mm^2/s}$ and $U_0=2.30~\mathrm{m/s}$. The results show that $\beta$ increases with $Re_p$. Based on nearly inviscid experiments\cite{Kiyama2016}, we set $\beta_0=2.05$. At two extreme conditions, $\left.\beta\right|_{Re_p=\infty}=\beta_0$ and $\left.\beta\right|_{Re_p=0}=0$. 

A damping prefactor, $(1-e^{-\zeta Re_p})$, is introduced to quantify the viscous effects,
\begin{equation}
    \beta = (1-e^{-\zeta Re_p})\beta_0.
    \label{Eq.beta}
\end{equation}
By fitting all the experimental and numerical results, we obtain $\zeta=0.076$. The prediction curve in Fig.~\ref{Fig.jetv} provides the best fit.
Specially, it is observed that the viscous effects can be neglected when $Re_p \geq 80$.

To understand the impact of surface tension on $\zeta$, we conducted additional numerical simulations where the surface tension coefficient was varied to 10 and $40~\mathrm{mN/m}$. The liquid viscosity was maintained at $22~\mathrm{mm^2/s}$, while $U_0$ ranged from 0.27 to $4.95~\mathrm{m/s}$. As shown in Fig.~\ref{Fig.jetv}, the data points corresponding to different surface tensions exhibit minimal deviation from the proposed relationship depicted in Eq.~\ref{Eq.beta}. The results confirmed that $\zeta$ is unaffected by changes in surface tension.

Furthermore, our study shows that $\zeta$ is also independent of viscosity. $\zeta$ quantifies the relationship between viscous energy dissipation and the Reynolds number $Re_p$. During a jet's development, its velocity continuously changes, resulting in ongoing viscous energy loss. This viscous energy loss becomes minor until the jet is fully developed. Establishing a reliable $\zeta$ requires the jet to be fully developed. In our study, we used a jet development time $\Delta t = 6$~ms to ensure the jets were fully developed. 
We tested a broad range of viscosities, from 0.5 to 5888~$\mathrm{mm^2/s}$, and surface tensions of 10, 20, and 40 mN/m. By fitting $\zeta$ with varying data ranges, we observed that $\zeta$ exhibited little variation. This consistency suggests that $\zeta$ can be considered constant, regardless of changes in viscosity and surface tension.

In summary, we investigate the viscous influences on impulsively generated focused jets. 
Experiments are performed with three kinds of viscous liquids, along with numerical simulations to gain more flow details during jet formation.
The results show that the dimensionless jet velocity decreases with an increase in liquid viscosity or a decrease in impact velocity. 
We demonstrate significant tangential flow occurs along the free surface, and the transfer of mass and momentum along the tangential direction of the surface contributes to the jet formation. 
A boundary layer is formed along the free surface, and the viscosity-induced diffusion of the shear flow and vorticity explains the reduction in jet speed. 
We propose an empirical equation to predict the viscous jet velocity after determining the boundary layer thickness. 
This study reveals the roles of the shear flow along the free surface and the corresponding viscous effects on impulsively generated focused jets. 
These findings provide new perspectives on viscous-interface dynamics in advanced manufacturing and biomedical applications.

\begin{acknowledgments}X. Cheng acknowledges Dr. Jiaojiao Guo at Wuhan University for helping with numerical simulations. This work was supported by the National Natural Science Foundation of China (grant no. 11872315), the Guangdong Basic and Applied Basic Research Foundation (grant no. 2022A1515011201), and the Innovation Capability Support Program of Shaanxi (program no. 2024RS-CXTD-15). L. J. acknowledges the support from the Royal Society Research Fund (RGS\textbackslash R2\textbackslash 222218).
\end{acknowledgments}

\end{document}